\documentclass{article}
\usepackage{spconf,amsmath,graphicx}

\usepackage{amssymb}
\usepackage{mathrsfs}
\usepackage{verbatim}
\usepackage{graphicx}
\usepackage{epstopdf}
\usepackage{cite}
\DeclareMathOperator{\sinc}{sinc} 
\DeclareMathOperator{\rect}{rect}

\title{Matched Illumination Waveforms using Multi-Tone Sinusoidal Frequency Modulation}
\twoauthors
{Kaushallya Adhikari\sthanks{Funded by Office of Naval Research (ONR) Summer Faculty Research Program.  Email: kadhikari@uri.edu}}
{Univerisity of Rhode Island\\
Electrical and Computer Engineering\\
45 Upper College Rd.,\\Kingston, RI, USA 02881}
{David A. Hague\sthanks{Funded by the Naval Undersea Warfare Center's In-House Laboratory Independent Research (ILIR) program.  Email: david.a.hague@ieee.org}}
{Naval Undersea Warfare Center\\
Sensors and Sonar Systems Department\\
1176 Howell St.,\\ Newport, RI, USA 02841}
\begin{document}
%
\maketitle
\begin{abstract}
This paper explores the design of constant modulus Matched-Illumination (MI) waveforms using the Multi-Tone Sinusoidal Frequency Modulation (MTSFM) waveform model.  MI waveforms are optimized for detecting targets in known noise and clutter Power Spectral Densities (PSDs).  There exist well-defined information theoretic methods that describe the design of MI waveforms for a myriad of target/noise/clutter models.  However, these methods generally only produce the magnitude square of the MI waveform's spectrum. Additionally, the waveform's time-series is not guaranteed to be constant modulus. The MTSFM is a constant modulus waveform model with a discrete set of design coefficients. The coefficients are adjusted to synthesize constant modulus waveforms that approximate the ideal MI waveform's spectrum.  Simulations demonstrate that the MTSFM's detection performance closely approximates an ideal MI waveform spectrum and generally outperforms flat spectrum waveforms across a range of transmit energies when the noise and clutter PSDs vary greatly across the operational band. 

\end{abstract}
\section{Introduction}
\label{sec:intro}
A fundamental problem in radar system design is the choice of transmit waveform and receiver processing for detecting targets of interest.  There exist well-defined information theoretic methods in the published literature that describe how to design waveforms and receivers for optimal target detection in given noise/clutter Power Spectral Densties (PSDs) \cite{MRBell_Info, Romero, pillai2000optimal, Kay2007, pillai1999, Guerci2000}.  Using a transmit energy constraint, these methods detail the structure of the magnitude square of the waveform's spectrum, also known as the Energy Spectral Density (ESD).  Such waveform design methods have demonstrated clear improvement in detection performance compared to waveforms with a flat ESD such as the Linear Frequency Modulated (LFM) waveform \cite{Kay2007, pillai1999}.  The improved detection performance of these Matched Illumination (MI) waveforms over flat spectrum waveforms is especially noticeable in scenarios where there is limited available transmit energy and the noise/clutter PSDs vary widely in magnitude across the operational band of frequencies \cite{pillai1999}.

While these MI waveform design techniques specify the optimal waveform's ESD shape, they do not directly specify how to synthesize the waveform time-series that realizes that ESD shape.  Since the optimization is over a finite band, the resulting waveform time-series cannot be perfectly time-limited.  Additionally, it is generally desirable for waveforms to possess a constant modulus to facilitate transmission on practical transmitter electronics, a property that information theoretic MI waveform synthesis methods do not guarantee.  There have been a number of efforts to develop phase-retrieval algorithms that synthesize a constant modulus waveform whose ESD closely approximates the ideal MI waveform's ESD \cite{phaseRetrieval, pillai2009reconstruction, Goodman}.  

Recently, the Multi-Tone Sinusoidal Frequency Modulated (MTSFM) waveform was developed for use as an adaptive FM waveform model for cognitive radar and sonar systems \cite{Hague_MTSFM}.  The MTSFM is a constant modulus waveform model with a discrete set of parameters that can be adjusted to synthesize novel waveforms with desirable characteristics.   Previous work in \cite{Hague_MTSFM, Hague_EOA} demonstrated that the MTSFM's design coefficients can be finely tuned to produce waveforms with specific Ambiguity Function (AF) and Auto-Correlation Function (ACF) properties, a common focus area for adaptive waveform design \cite{AubryI, PalomarI, PalomarII, SoltanalianIII}.  This paper explores applying the MTSFM waveform model to the MI waveform design problem using the foundational methods developed by Kay in \cite{Kay2007} for point-like targets.  Simulations demonstrate that the MTSFM's detection performance approaches that of the ideal MI waveform and generally outperforms flat spectrum waveforms across a range of transmit energies when the magnitudes of the noise and clutter PSDs vary substantially across the operational band of frequencies.  

\section{Waveform Signal Model and the Optimal Detection Waveform Problem}
\label{sec:sigModel}

This section describes the MI waveform design technique for point targets in noise and clutter whose respective PSDs are known \cite{Kay2007}.  This section additionally describes the MTSFM waveform model and how it can be adapted to produce constant amplitude waveforms that approximate the ideal MI waveform's ESD.  This paper assumes the waveform $s\left(t\right)$ with Fourier transform $S\left(f\right)$ is basebanded and occupies a bandwidth $W$.  The waveform is defined over the time interval $-T/2 \leq t \leq T/2$ with duration $T$ and energy $E$ expressed as 
\begin{equation}
s\left(t\right) = \sqrt{\dfrac{E}{T}}\rect\left(t/T\right)e^{j\varphi\left(t\right)}
\label{eq:complexExpo}
\end{equation}
where $\varphi\left(t\right)$ is the waveform's instantaneous phase.  The waveform's frequency modulation function $m\left(t\right)$ is expressed as 
\begin{equation}
m\left(t\right) = \dfrac{1}{2\pi}\dfrac{d\varphi\left(t\right)}{dt}.
\label{eq:mod}
\end{equation}

\subsection{Designing Matched Illumination Waveforms}
\label{subsec:optimalDetectionProblem}
\begin{figure}[ht]
\centering
\includegraphics[width=0.5\textwidth]{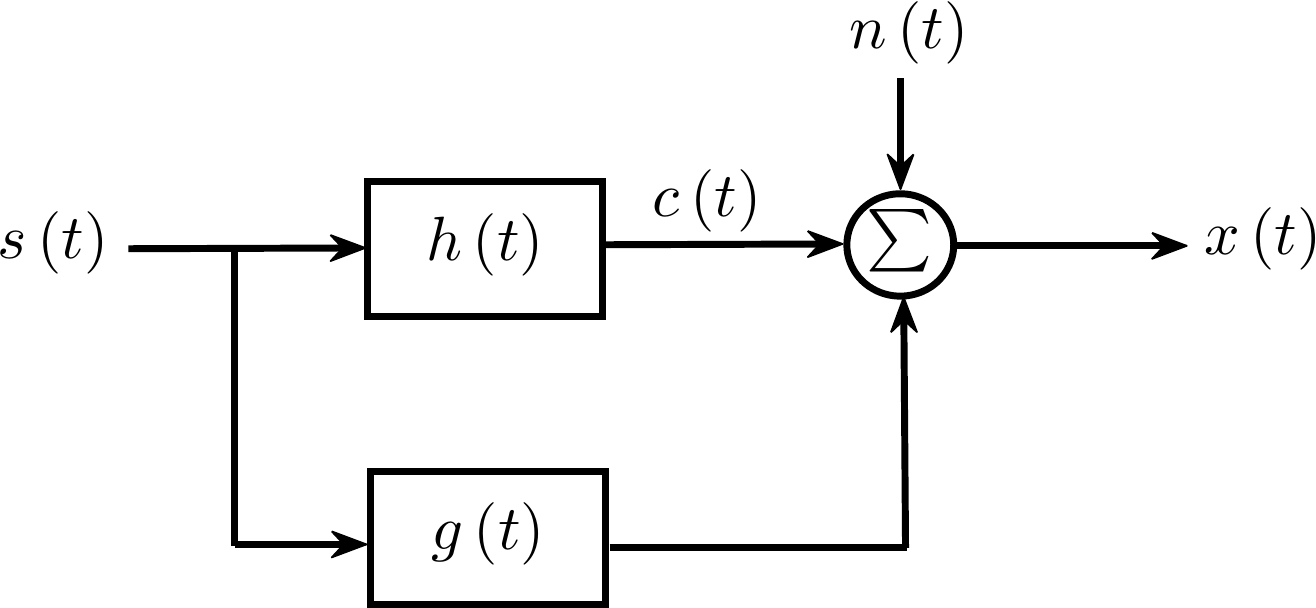}
\caption{Block diagram describing the target scene model.  The received signal at the target's time-delay of arrival is a superposition of a scaled version of the transmitted waveform $s\left(t\right)$ plus additive noise $n\left(t\right)$ and clutter $c\left(t\right)$ from the target scene.}
\label{fig:detectionProblem}
\end{figure}

This paper uses the MI waveform model developed by Kay in \cite{Kay2007} and is described by the block diagram shown in Figure \ref{fig:detectionProblem}.  The waveform $s\left(t\right)$ is transmitted into the medium.  The return signal is a combination of the return from the target, clutter, and additive noise.  The target is modeled as a point reflector with impulse response $g\left(t\right) = A \delta\left(t\right)$ where $\delta \left(t\right)$ is an impulse function and $A$ is a complex reflecting parameter modeled as a complex normal distribution $A\sim \mathcal{CN}\left(0, \sigma_A^2\right)$.  This target model can be readily generalized to include extended targets as was demonstrated in \cite{Romero}.  The noise $n\left(t\right)$ is modeled as a complex Gaussian random process with PSD $P_n\left(f\right)$.  The channel impulse response $h\left(t\right)$ is convolved with the transmit waveform producing the clutter signal $c\left(t\right) = h\left(t\right)*s\left(t\right)$.  Assuming the PSD of the channel is a Gaussian random process with zero mean and PSD $P_h\left(f\right)$, the PSD of the clutter can correspondingly be expressed as $P_c\left(f\right) = |S\left(f\right)|^2P_h\left(f\right)$.  As shown in Figure \ref{fig:detectionProblem}, these terms combine to produce the return signal $x\left(t\right)$ expressed as
\begin{equation}
x\left(t\right) = As\left(t\right) + h\left(t\right)*s\left(t\right) + n\left(t\right).
\end{equation}
Note that this model assumes the target and clutter are stationary and thus this signal model does not contain Doppler shifted echo signals or clutter.  This was primarily utilized in \cite{Kay2007} for simplicity in deriving the optimal waveform/receiver configuration.  However, this model also represents a worst case scenario.  Many radar/sonar systems exploit target Doppler in order to separate the target's echo signal from clutter.  For stationary targets and clutter this is not possible, and thus target detection performance is purely dependant upon receiver design and shaping of the waveform's ESD.

Kay \cite{Kay2007} then used this model to derive the optimal Neyman-Pearson detector, expressed in the frequency domain as
\begin{equation}
\left|\sum_{-M/2}^{M/2}\dfrac{X\left(f_m\right)S^*\left(f_m\right)}{P_h\left(f_m\right)|S\left(f_m\right)|^2 + P_n\left(f_m\right)}  \right|^2 > \gamma
\label{eq:receiver}
\end{equation}
where $f_m = m/T$, $M = \lceil WT \rceil$, and $\gamma$ is the detection threshold.  The optimal receiver's detection performance is determined by the metric
\begin{equation}
d^2 = \sigma_A^2\int_{-W/2}^{W/2}\dfrac{|S\left(f\right)|^2}{P_h\left(f\right) |S\left(f\right)|^2 + P_n\left(f\right)}df.
\label{eq:detectionMetric}
\end{equation}
The waveform is also constrained to possess finite energy across the operational band of frequencies $W$
\begin{equation}
E=\int_{W}|S\left(f\right)|^2df.
\label{eq:Econ}
\end{equation}
The waveform that maximizes $d^2$ possesses the ESD
\begin{equation}
E_s\left(f\right) = |S\left(f\right)|^2 =  \max\left(\dfrac{\sqrt{P_n\left(f\right)/\lambda}-P_n\left(f\right)}{P_h\left(f\right)}, 0\right)
\label{eq:OptimalESD}
\end{equation}
where the parameter $\lambda$ is found from the energy constraint in \eqref{eq:Econ}.  This involves solving the following expression
\begin{equation}
\int_{W}\max\left(\dfrac{\sqrt{P_n\left(f\right)/\lambda}-P_n\left(f\right)}{P_h\left(f\right)}, 0\right)df = E.
\label{eq:eConstraint}
\end{equation}
The value for $\lambda$ can be solved for numerically given $P_n\left(f\right)$ and $P_h\left(f\right)$.  Using the receiver in \eqref{eq:detectionMetric}, the Receiver Operating Characteristic (ROC) which relates the probability of detection $P_D$ and the probability of false alarm $P_{FA}$ is completely characterized using the detection metric defined in \eqref{eq:detectionMetric} and is expressed as \cite{Kay2007}
\begin{equation}
P_D = P_{FA}^{\frac{1}{1 + d^2}}.
\label{eq:ROC}
\end{equation}
Setting $d^2 = 0$ results in the line of no discrimination ROC curve where $P_D = P_{FA}$.  As $d^2 \to\infty$, the ROC curve approaches the perfect detector.  Maximizing the detection metric $d^2$ via the MI waveform described by \eqref{eq:OptimalESD} and \eqref{eq:eConstraint} and its corresponding receiver \eqref{eq:receiver} will therefore maximize the detection probability $P_D$ for a given fixed false alarm probability $P_{FA}$.  Thus, the detection metric $d^2$ in \eqref{eq:detectionMetric} is the primary figure of merit this paper uses to evaluate MI waveform designs using the model developed by Kay \cite{Kay2007}.

\subsection{The MTSFM Waveform Model}
\label{subsec:MTSFM}

The MTSFM waveform is realized by representing the modulation function as a finite Fourier series expansion.  While this paper focuses on modulation functions composed solely of sine harmonics for simplicity, the model can be readily generalized to include cosine harmonics as well.  The MTSFM's modulation function is expressed as
\begin{equation}
m\left(t\right) = \sum_{k=1}^K b_k \sin\left(\frac{2 \pi k t}{T}\right).
\label{eq:MTSFM_mod}
\end{equation}
The corresponding phase modulation function is expressed as
\begin{equation}
\varphi\left(t\right) = -\sum_{k=1}^K \beta_k \cos\left(\frac{2 \pi k t}{T}\right), \label{eq:MTSFM_phase}
\end{equation}
where $\beta_k=\left(\frac{b_k T}{k}\right)$ are the waveform's modulation indices and serve as a discrete set of $K$ parameters that adapts the waveform's characteristics.   Inserting \eqref{eq:MTSFM_phase} into  \eqref{eq:complexExpo} yields the MTSFM waveform's time-series.  

The MTSFM waveform time-series can also be represented as a complex Fourier series expressed as \cite{Hague_MTSFM}
\begin{equation}
s\left(t\right) = \sqrt{\frac{E}{T}}\rect\left(t/T\right)\sum_{m=-\infty}^{\infty} c_m e^{j\frac{2\pi m t}{T}}
\label{eq:mtsfmGBF}
\end{equation}
where the Fourier series coefficients $c_m$ are the Modified Generalized Bessel Functions (M-GBFs) \cite{DattoliBook} with integer order $m$ expressed as $\mathcal{I}_m^{1:K}\left(\{-j\beta_k\}\right)$.  The expansion in \eqref{eq:mtsfmGBF} shows that the MTSFM belongs to the family of generalized multi-carrier waveform models such as Orthogonal Frequency Division Multiplexing (OFDM) \cite{Koivunen}.  A unique characteristic of the MTSFM model is that unlike standard OFDM models with a generic set of coefficients $c_m$, the GBF coefficients of the MTSFM model ensures the resulting waveform is constant modulus \cite{Hague_MTSFM, HagueASA}.  The spectrum of the MTSFM waveform is expressed as an orthonormal superposition of frequency shifted $\sinc$ functions \cite{Hague_MTSFM}
\begin{equation}
S\left(f\right) = \sqrt{ET} \sum_{m=-\infty}^{\infty}\mathcal{I}_m^{1:K}\left(\{-j\beta_k\}\right) \sinc\left[\pi T \left(f-\frac{m}{T}\right)\right].
\label{eq:MTSFM_Spec}
\end{equation}
The design goal is to now use the MTSFM waveform model to approximate the optimal ESD of the MI waveform design problem specified by \eqref{eq:OptimalESD} and \eqref{eq:eConstraint}.

\subsection{The Design of MI Waveforms using the MTSFM Model}
\label{subsec:MI_MTSFM}
This section describes a heuristic structured phase retrieval method to design a constant amplitude MTSFM waveform whose ESD $|S\left(f\right)|^2$ closely approximates the ESD of the MI waveform defined in Section \ref{subsec:optimalDetectionProblem} denoted as $|S_o\left(f\right)|^2$.  The first step is to find $|S_o\left(f\right)|^2$ using \eqref{eq:OptimalESD} and \eqref{eq:eConstraint}.  Since the MI waveform design method is concerned only with the ESD, the phase of the MI waveform's spectrum can be ignored and therefore $S_o\left(f\right) = |S_o\left(f\right)|$.  Discretizing $S_o\left(f\right)$, the generic OFDM coefficients $c_m$ can be solved in matrix form via $\underline{\mathbf{s_o}}=\mathbf{X}\mathbf{\underline{c}}$.  Here, $\underline{\mathbf{s_o}}$ is the discrete vector form of $S_o\left(f\right)$, the vector $\underline{\mathbf{c}}$ represents the generic OFDM coefficients $c_m$, and $\mathbf{X}$ is a matrix composed of discretized frequency shifted versions of the $\sinc$ function in \eqref{eq:MTSFM_Spec} with frequency spacing $f_m = m/T$ as in \eqref{eq:receiver}.  The frequency spacing results in $\mathbf{X}$ being square and invertible.  Thus a unique solution to $c_m$ exists by solving $\underline{\mathbf{s_o}}=\mathbf{X}\mathbf{\underline{c}}$ via $\underline{\mathbf{c}}=\mathbf{X}^{-1}\underline{\mathbf{s_o}}$.

The resulting coefficients $c_m$ are generic OFDM coefficients and not guaranteed to synthesize a constant modulus waveform.  Therefore, synthesizing a constant modulus MTSFM waveform that approximates $|S_o\left(f\right)|$ requires finding a M-GBF based fit to the coefficients $c_m$.  While the coefficients $c_m$ are real but not necessarily positive, the M-GBF coefficients can be complex valued.  Thus, the authors propose synthesizing a MTSFM approximation to the ideal MI waveform by minimizing the following distance metric between the magnitudes $|c_m|^2 = c_m^2$ and $|\mathcal{I}_m^{1:K}\left(\{-j\beta_k\}\right)|^2$ 
\begin{multline}
\underset{\beta_k}{\text{min}}\text{~}F\left(\{\beta_k\}\right) = \| c_m^2 - E|\mathcal{I}_m^{1:K}\left(\{-j\beta_k\}\right)|^2 \|_2^2 \\ \text{~s.t.} \sum_k k\beta_k \in \left(1\pm\delta\right)\kappa.
\label{eq:LS_1}
\end{multline}
where $\kappa$ is the region of support of $c_m$ such that $\sum_{m\in \kappa} c_m^2 \cong E$ and the $\sum_k k \beta_k$ term loosely approximates the M-GBF's region of support \cite{DattoliBook}.  Note that the $E$ term in front of the M-GBF argument ensures proper scaling with $c_m^2$ since $\sum_m c_m^2 = E$ and $\sum_m |\mathcal{I}_m^{1:K}\left(\{-j\beta_k\}\right)|^2 = 1$ \cite{DattoliBook}.  The quartic objective function defined in \eqref{eq:LS_1} is loosely similar to those defined in other generalized phase retrieval problems \cite{sun2018geometric, PR_Wirtinger}.  The quartic nature of this distance metric between $c_m^2$ and $|\mathcal{I}_m^{1:K}\left(\{-j\beta_k\}\right)|^2$ coupled with the highly oscillatory nature of the M-GBFs with varying arguments make the objective function defined in \eqref{eq:LS_1} nonconvex.  Such an objective function makes it unlikely that common iterative methods will solve this problem without special consideration to initialization.  However, efforts in the literature \cite{nonConvexPhase, PR_Optics} have demonstrated that heuristic methods work surprisingly well on these nonconvex problems and produce useful results.  

\section{Two Illustrative Design Examples}
\label{sec:examples}
\begin{figure}[ht]
\centering
\includegraphics[width=0.5\textwidth]{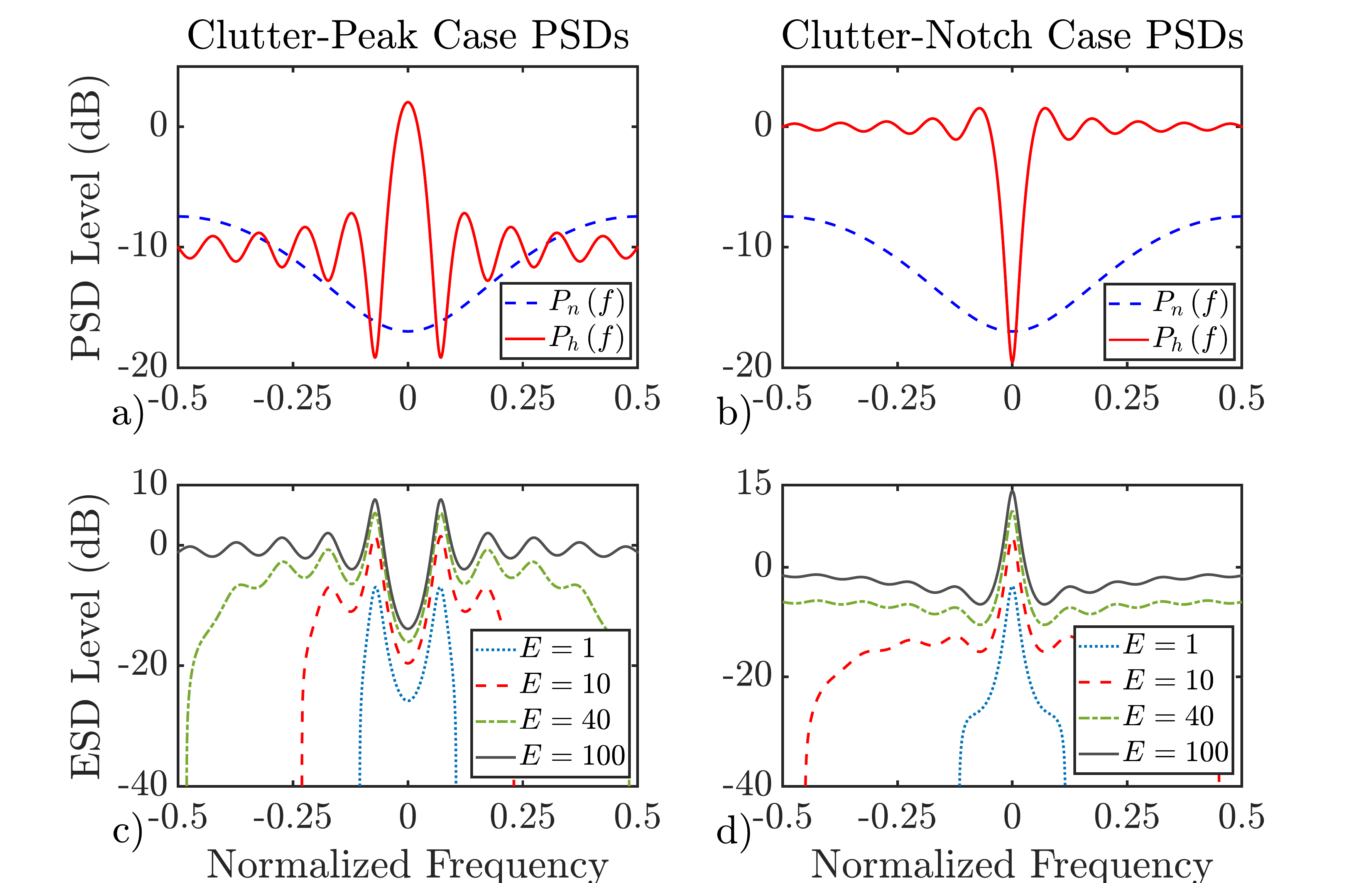}
\caption{Illustration of the noise/clutter PSDs and their corresponding MI waveform ESDs for several energy values.}
\label{fig:clutterCases}
\end{figure}

The following simulations demonstrate the MTSFM-based fit to the MI waveform problem.  Figure \ref{fig:clutterCases} shows the noise and clutter PSDs for two scenarios as well as the ideal MI waveforms for several energy values.  These simlulations utilize the same point target statistics ($\sigma_A^2 = 1$) and noise PSDs while using different clutter PSDs.  The noise PSD has a relatively broad valley centered about DC.  Noise only water-filling techniques commonly utilized in MI waveform design would therefore emphasize most of the waveform's energy about DC.  However, the clutter also heavily influences the MI waveform's ESD.  The first scenario's clutter PSD is oscillatory across most of the operational band with a distinct peak at DC (i.e, the ``clutter-peak'' case).  The second scenario's clutter PSD is largely flat across the operational band with a distinct notch centered about DC (i.e, the ``clutter-notch'' case).  Using \eqref{eq:OptimalESD} and \eqref{eq:eConstraint} produces MI waveforms whose ESDs vary roughly 20 dB across the operational band. 

\begin{figure}[ht]
\centering
\includegraphics[width=0.5\textwidth]{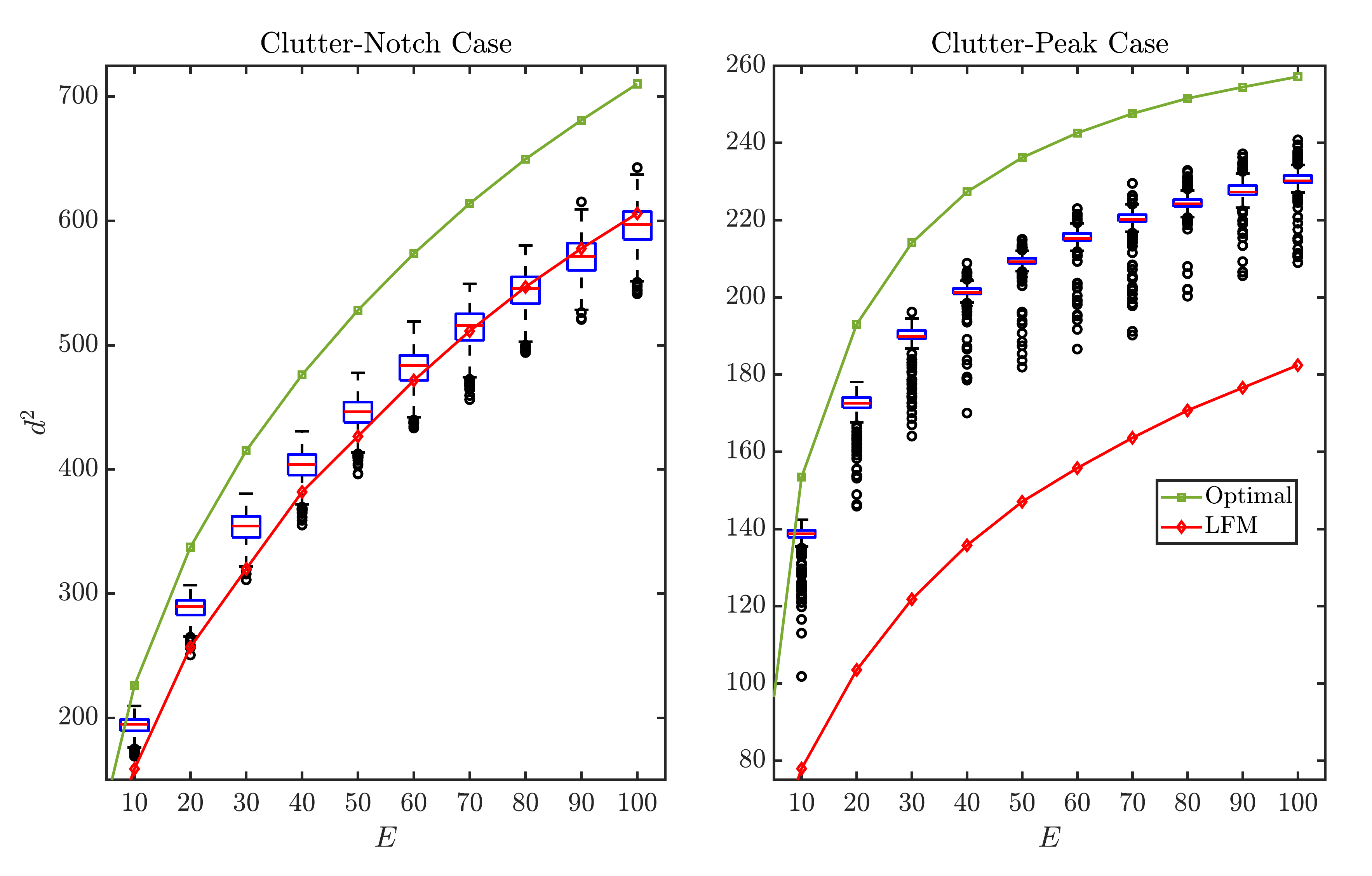}
\caption{Box-whisker plots of the detection metric $d^2$ of 1000 MTSFM trials for each energy value and the detection metrics for both the ideal MI waveforms and LFM waveforms with equivalent RMS bandwidth for each energy value.}
\label{fig:exampleI}
\end{figure}

Figure \ref{fig:exampleI} shows box-whisker plots of the detection metric $d^2$ of the MTSFMs fitted to the ideal MI waveform ESDs across a range of energy values for both scenarios shown in Figure \ref{fig:clutterCases}.  For each energy value, 1000 MTSFM waveforms each with a different set of initial modulation indices $\beta_k$ were fit to the ideal MI waveform's ESD using \eqref{eq:LS_1} with $\delta = 0.2$.  The black circles denote statistical outliers.  Addtionally, the detection metric for an LFM waveform with equal RMS bandwidth $\beta_{rms}^2 = \left(2\pi\right)^2/E\int_{W} f^2|S\left(f\right)|^2 df$ to that of the MI and MTSFM waveforms is also shown for each energy value.  For the clutter-peak case, the MTSFMs outperform the LFM often for lower energies but noticeably less so for higher energies.  One potential explanation for this result is that for higher energies, the MI waveform starts to resemble a flat spectrum waveform.  An analysis of the trials showed that the MTSFM tends to better fit spectral shapes where there is notable variation across the operational band.  For the clutter-notch case every MTSFM outperformed the LFM for energy values $E > 1$.  This is likely because the MI waveform's ESD shapes possess a distinct peak at DC, a spectral shape the MTSFM is much better suited to fitting than the flat spectrum LFM.

\section{Conclusion}
\label{sec:Conclusion}
This paper explored applying the MTSFM waveform model to the MI waveform design problem for point-like targets using the model in \cite{Kay2007}.  The M-GBF coefficients that describe the MTSFM are fit to a set of ideal OFDM coefficients $c_m$ via the nonconvex distance metric in \eqref{eq:LS_1}.  Simulations show that the MTSFM on average produces waveforms whose detection performance tends to exceed that of spectrally flat waveforms when the ideal MI waveform's ESD varies substantially across the operational band.  There are several future avenues to pursue with this work.  The most obvious is expanding this analysis to extended targets with MTSFM waveforms whose modulation functions include cosine and sine harmonics which produce a richer set of realizable spectral shapes.  Another avenue is refining the phase retrieval process using methods from \cite{nonConvexPhase} to produce MTSFM waveforms whose ESDs more tightly fit the ideal MI waveform's ESD.  


\clearpage

\bibliographystyle{IEEEbib}

\end{document}